\documentstyle[12pt]{article}
\renewcommand{\theequation}{\arabic{equation}}
\newcommand{\EQ}{\begin{equation}}
\newcommand{\EN}{\end{equation}}

\newcommand{\bear}{\begin{eqnarray}}
\newcommand{\ear}{\end{eqnarray}}

\begin{document}

\topmargin 0pt
\oddsidemargin 5mm
\newcommand{\NP}[1]{Nucl.\ Phys.\ {\bf #1}}
\newcommand{\PL}[1]{Phys.\ Lett.\ {\bf #1}}
\newcommand{\NC}[1]{Nuovo Cimento {\bf #1}}
\newcommand{\CMP}[1]{Comm.\ Math.\ Phys.\ {\bf #1}}
\newcommand{\PR}[1]{Phys.\ Rev.\ {\bf #1}}
\newcommand{\PRL}[1]{Phys.\ Rev.\ Lett.\ {\bf #1}}
\newcommand{\MPL}[1]{Mod.\ Phys.\ Lett.\ {\bf #1}}
\newcommand{\JETP}[1]{Sov.\ Phys.\ JETP {\bf #1}}
\newcommand{\TMP}[1]{Teor.\ Mat.\ Fiz.\ {\bf #1}}

\renewcommand{\thefootnote}{\fnsymbol{footnote}}

\newpage
\setcounter{page}{0}
\begin{titlepage}
\begin{flushright}
UFSCARF-TH-94-22
\end{flushright}
\vspace{0.5cm}
\begin{center}
{\large The exact solution and the finite size behaviour of the $Osp(1|2)$
invariant spin chain}\\
\vspace{1cm}
\vspace{1cm}
{\large M.J.  Martins  } \\
\vspace{1cm}
{\em Universidade Federal de S\~ao Carlos\\
Departamento de F\'isica \\
C.P. 676, 13560~~S\~ao Carlos, Brasil}\\
\end{center}
\vspace{1.2cm}

\begin{abstract}
We have solved exactly the $Osp(1|2)$ spin chain by the Bethe ansatz approach.
Our solution is based on an
equivalence between the $Osp(1|2)$ chain and certain special limit of the
Izergin-Korepin
vertex model. The completeness of
the Bethe ansatz equations is discussed for a system with four sites and it is
noted the appearance of
special string structures. The Bethe ansatz presents an important phase-factor
which distinguishes the
even and odd sectors of the theory. The finite size properties are governed by
a conformal field theory
with central charge $c=1$.
\end{abstract}
\vspace{.2cm}
\vspace{.2cm}
\centerline{December 1994}
\end{titlepage}

\renewcommand{\thefootnote}{\arabic{footnote}}
\setcounter{footnote}{0}

\newpage
\section{Introduction}
Exactly integrable vertex models possessing both fermionic and bosonic degrees
of freedom provide
interesting examples of interacting systems presenting a very rich structure in
their spectrum. The exact
solution of Hamiltonians which appears as graded permutations of bosons and
fermions
was first considered by Sutherland \cite{SUT}.
In general, the associated Boltzmann weights satisfy a graded version \cite{KS}
of the Yang-Baxter
equation and they have been investigated as invariants under the superalgebras
$Sl(n|m)$ and $Osp(n|2m)$
\cite {KUU,BA}.
In particular, the basis of the quantum inverse scattering method for certain
superorthosymplectic magnets has been developed by Kulish \cite{KUU}.
An important example is the Perk-Schultz system \cite {CLS,PK,DV}, which
recently has been
recognized to appear on the solution of several models of correlated electrons
on a lattice \cite{KO1,KO2}.
For instance, the solution of the one dimensional supersymmetric $t-J$ model
\cite{TJ} is related
to the Bethe ansatz properties of the $Sl(1|2)$ invariant Perk-Schultz like
model \cite {CLS,KO1,KF}. In this
sense, it seems quite important to search for solutions of other integrable
systems possessing
bosonic/fermionic degrees of freedom.

In this paper, we focus on the exact solution of the simplest
super-orthosymplectic invariant spin chain.
Their Boltzmann weight has three states per bond, one bosonic and two
fermionic, and the associated
spin magnet is invariant under $Osp(1|2)$ symmetry \cite{KUU}. Its
corresponding $R$-matrix \cite{KUU} is given
by ( see also \cite{BA,MA,CH} )
\EQ
R(\lambda, \eta)_{i,i+1} = \lambda I_{i,i+1} +\eta P^g_{i,i+1}
+\frac{\eta \lambda}{3\eta/2 -\lambda}E^g_{i,i+1}
\EN
where $\lambda$ is the spectral variable, $\eta$ is the quasi-classical
parameter and $I_{i,i+1}$ is
the $9 \times 9$ identity matrix. The graded permutation operator $P^g$ has the
elements $(P^g_{i,i+1})_{ab}^{cd}=
(-1)^{p(a)p(b)} \delta_{a,d} \delta_{b,c}$ where $p(1)=0$ (boson),
$p(2)=p(3)=1$ (fermions) in the order of BFF
grading \cite{KO1}. $(E^g_{i,i+1})_{ab}^{cd}= \alpha_{ab} \alpha_{cd}^{st}$ is
the $Osp(1|2)$
Temperely-Lieb operator \cite {MA,CH} and the symbol $st$ indicates the
supertranspose operation. The
matrix $\alpha$ on the specific $BFF$ grading has the following form
\EQ
\alpha=\left( \begin{array}{ccc}
	1 &   0 & 0 \\
	0 &  0& 1 \\
	0 & -1 & 0 \\
	\end{array}
	\right)
\EN

The $R$-matrix (1) has the important property of being proportional to the
graded operator $P^g$ at
the special point $\lambda =0$. As a consequence, the local Hamiltonian is
obtained as a logarithmic
derivative of the Transfer matrix at point $\lambda =0$ \cite {KO1}. The
associated Hamiltonian is then
given by
\EQ
H= -J\sum_{i=1}^{L} [ P^g_{i,i+1} +\frac{2}{3} E^g_{i,i+1} ]
\EN
where periodic boundary condition is implicitly assumed. The antiferromagnetic
regime of (3)
corresponds to $J >0$. The $Osp(1|2)$ invariance of this system is shown in
Appendix A.

In this paper we use the analytical/algebraic Bethe ansatz \cite{RE,TA} in
order to compute the eigenvalues
of the Hamiltonian (3). By means of a canonical transformation we are able to
write (1) as a certain
limit of the vertex operator of the Izergin-Korepin (IK) model \cite{IK}. The
associated Bethe ansatz
equation has a peculiar phase behaviour, which is important for the correct
characterization of the
critical properties.

This paper is organized as follows\footnote{A brief account of our results has
appeared in ref. \cite{MAR}}. In section 2, we present the Bethe ansatz
equations associated to
the diagonalization of the
Hamiltonian  (3). In section (3), we discuss the completeness of the Bethe
ansatz roots
for a lattice of size $ L=4$. In particular we present  evidence that special
string structures may
appear in the spectrum. The thermodynamic limit and the critical behaviour is
computed in section 4.
In section 5 we summarize our conclusions and discuss some remaining questions.
In Appendix A and B we
show the $Osp(1|2)$ invariance of the Hamiltonian (3) as well as its relation
to the IK model, respectively.
In Appendix $C$ some results for twisted boundary conditions are discussed.
\section{The Bethe ansatz solution}

In this section we are going to argue that
the problem of diagonalization of the Hamiltonian (3) is similar to that
performed in the IK vertex model \cite{RE,TA}.
We recall that   more general treatments
of the IK vertex model based on the  $A_2^2 $ algebra and its RSOS reductions
can
be found,  for instance, in refs. \cite{JI,JI1}.
In order to see this equivalence, it is convenient to work with a vertex
operator $ {\cal L}(\lambda,\eta) $ satisfying the usual
Yang-Baxter equation. As it has been first discussed in ref. \cite{KS}, this
can be done
(without changing the original problem) because the $R$-matrix (1) is shown
to be obtained \cite{MA} from a null-parity (Grassmann) braid operator. Their
relation is rather simple \cite{KS}
\EQ
{\cal L}_{ab}^{cd} (\lambda, \eta) = (-1)^{p(a)p(b)} R_{ab}^{cd}(\lambda,\eta)
\EN

The next step corresponds to making a redefinition of the grading to FBF, and
afterwords
 to rewrite this operator
in terms
of the $SU(2)$  spin-1 generators. By performing
this canonical transformation, the operator ${\cal L}(\lambda,\eta)$ is given
in terms of spin-1 matrices
$S^{\pm}$, $S^z$ by the following expression
\EQ
{\cal L}(\lambda,\eta)= \left( \begin{array}{ccc}
          \begin{array}{c} \lambda I +f(\lambda,\eta) S^{z} - \\
{\tilde{g}}(\lambda,\eta)(S^{z})^{2} \end{array} & \frac{1}{\sqrt{2}}[\eta S^-
S^z
-2\lambda h(\lambda,\eta) S^{z} S^{-}] & f(\lambda,\eta) (S^-)^2 \\
           \begin{array}{c} \frac{1}{\sqrt{2}}[\eta S^z S^+ -\\
2\lambda h(\lambda,\eta) S^{+} S^{z}]\end{array} & \begin{array}{c} [\lambda -3
\eta h(\lambda,\eta)]I \\
+3 \eta h(\lambda,\eta)  (S^z)^2 \end{array} &
          \begin{array}{c} -\frac{1}{\sqrt{2}}[\eta S^z S^- + \\
2\lambda h(\lambda,\eta) S^{-} S^{z}] \end{array} \\
 f(\lambda,\eta) (S^+)^2 &
           -\frac{1}{\sqrt{2}}[\eta S^+ S^z
+2\lambda h(\lambda,\eta) S^{z} S^{+}]  &
          \begin{array}{c} \lambda I -f(\lambda,\eta) S^{z} \\
-{\tilde{g}}(\lambda,\eta)(S^{z})^{2} \end{array}\\
\end{array}
\right)
\EN
where $I$ is the  $3 \times 3$ identity matrix and functions $f(\lambda,\eta)$,
${\tilde{g}}(\lambda,\eta)$ and $h(\lambda,\eta)$ are
given by
\EQ
f(\lambda,\eta)=\eta \frac{(2\lambda-3\eta/2)}{2(\lambda-3\eta/2)};~~
h(\lambda,\eta)=\frac{\eta}{2(\lambda-3\eta/2)};~~{\tilde{g}}(\lambda,\eta)=
2\lambda +\frac{3\eta}{2}h(\lambda,\eta)
\EN

The Transfer matrix $T(\lambda,\eta)$ defined on the Hilbert space of $L$ sites
is the generator of commuting quantum integrals of motion. As usual,  it is
built up
in terms of the vertex operators by the expression
\EQ
T(\lambda,\eta) =Tr_0[ {\cal L}_{0L}(\lambda,\eta) \cdots {\cal
L}_{01}(\lambda,\eta)]
\EN
where the index $0$ stands for the $3 \times 3$ auxiliary space. The ${\cal
L}(\lambda,\eta)$-matrix at $\lambda=0$ is
proportional to the operator of permutations and the logarithmic derivative of
$T(\lambda,\eta)$ at this point defines
the corresponding spin-1 Hamiltonian. The associated spin-1 chain is given by
\bear
H= \sum_{i=1}^{L} \left[ \frac{1}{3} \sigma_i^2 -\sigma_i
-\frac{4}{3}[(\sigma_i^z)^2 - \sigma_i +
\sigma_i^z \sigma_i^{\perp} + \sigma_i^{\perp} \sigma_i^z] +2 [(S_i^z)^2
+(S_{i+1}^z)^2] -\frac{7}{3}I
\right . \nonumber \\ \left .
-\frac{1}{3}[\sigma_i^z(S_i^+ S_{i+1}^- -S_i^{-}S_{i+1}^{+}) +
(S_i^+ S_{i+1}^- -S_i^{-}S_{i+1}^{+}) \sigma_i^z]
\right ]
\ear
where $\sigma_i= S_i . S_{i+1}= \sigma_i^z +\sigma_i^{\perp}$ and $\sigma_i^z=
S_i^z S_{i+1}^z $.

The first step toward the diagonalization of (8) is to
 notice that the vertex operator ${\cal L}(\lambda,\eta)$
resembles much that appearing in
the construction of IK vertex model \cite{IK,JI,JI1}. First of all, it is
possible to verify that the operator
$ {\cal L}(\lambda,\eta)$ gives origin to 19 nonvanishing Boltzmann weights on
a square lattice.
This is schematized in Fig.1 by putting the values $\pm,0$ on each bond of the
lattice and by assuming a node
current conservation. This picture suggests that a connection with the IK
vertex model can be indeed tried. In fact,
in Appendix B, we show that the
Hamiltonian (8) can be obtained as an appropriate limit of that associated to
the IK model. We also discuss
how the algebraic Bethe ansatz
method developed by Tarasov \cite{TA} can be directly applied to this problem.
In the
following we present the main steps of
the analytical Bethe ansatz approach used in ref. \cite{RE}. We take as
reference state the ferromagnetic vacuum defined by
\EQ
|0> = \prod_{i}^{L} |0>_i; ~~~ |0>_i= \left( \begin{array}{c} 1 \\ 0 \\0
\end{array} \right )
\EN

The matrix $ {\cal L}(\lambda,\eta) $ acting on the reference state
has a triangular form with respect to the auxiliary
space, namely
\EQ
{\cal L}(\lambda,\eta) |0> = \left ( \begin{array}{ccc} \eta -\lambda & * & *
\\
0  & \lambda & * \\
0 & 0 & \frac{\lambda(\eta/2 -\lambda)}{\lambda -3 \eta/2} \end{array} \right )
\EN

As a consequence, the corresponding eigenvalue $ \Lambda(\lambda)$ of the
Transfer matrix acting on this state is
given by
\EQ
\Lambda(\lambda)= (\eta-\lambda)^L +\lambda^L +\left [
\frac{\lambda(\eta/2-\lambda)}{\lambda-3 \eta/2} \right ]^L
\EN

In order to construct other eigenvalues, the analytical approach seeks for a
more general ansatz of form
\EQ
\Lambda(\lambda,\{ \lambda_ j\}) =
(\eta-\lambda)^L \prod_{j=1}^{M} A(\lambda_j-\lambda) +\lambda^L
\prod_{j=1}^{M} B(\lambda_j-\lambda)
+\left [ \frac{\lambda(\eta/2-\lambda)}{\lambda-3 \eta/2} \right ]^L
\prod_{j=1}^{M} C(\lambda_j-\lambda)
\EN

Following the arguments of ref. \cite{RE}, crossing symmetry and unitarity
condition of the vertex ${\cal L}(\lambda,\eta)$
and some analytical hypotheses concerning the behaviour of the Transfer matrix
(7) fix functions $A(y)$, $B(y)$ and
$C(y)$ to be
\EQ
A(y)=\frac{\eta/2-y}{y+\eta/2},~~ C(y-3\eta/2) =A^{-1}(y),~~ B(y)=
A^{-1}(y+\eta/2)A(-y-\eta)
\EN

Collecting these results all together we finally find
\bear
\Lambda(\lambda,\{ \lambda_ j\}) =
(i-\lambda)^L \prod_{j=1}^{M}
-\frac{\lambda_j-\lambda-i/2}{\lambda_j-\lambda+i/2}
+\lambda^L \prod_{j=1}^{M} \frac{\lambda_j-\lambda }{\lambda_j-\lambda+i}
\frac{\lambda-\lambda_j-3i/2}{\lambda-\lambda_j-i/2}
\nonumber \\ +
\left [ \frac{\lambda(i/2-\lambda)}{\lambda-3 i/2} \right ]^L \prod_{j=1}^{M}
-\frac{\lambda-\lambda_j-2i}{\lambda-\lambda_j-i}
\ear
where due to the scale invariance
$ {\cal L}(\eta \lambda, \eta)= \eta {\cal L}(\lambda,1)$ we have chosen the
parameter $\eta =i$.

The set of numbers $\{\lambda_j \}$ are then fixed by imposing that
function $\Lambda(\lambda,\{\lambda_j\})$ has
no pole at finite value of $\lambda$. This means that the residues of
$\Lambda(\lambda,\{\lambda_j \}) $ at the
poles $\lambda= \lambda_j +i/2 $ and $\lambda= \lambda_j +i$ must vanish. An
important check of the ansatz (14)
is that these pole conditions should give the same restriction for the set $\{
\lambda_j \}$. In fact, this is
guaranteed by the crossing symmetry of the operator ${\cal L}(\lambda,\eta) $
and we find the following Bethe ansatz
equation
\EQ
{\left( \frac{\lambda_j -i/2}{\lambda_j +i/2} \right)}^{L}= - (-1)^r
\prod_{k=1}^{M}
\left(\frac{\lambda_j -\lambda_k -i}{\lambda_j-\lambda_k +i} \right)
\left(\frac{\lambda_j -\lambda_k +i/2}{\lambda_j-\lambda_k -i/2} \right)
\EN
where $r=L-M$ \footnote{
At this point we recall that in ref. \cite{KUU}, by using a different approach
of ours, the author presents a Bethe
ansatz equation without the phase factor $-(-1)^r$. We stress that his
discussion is very brief and at least for us
it is not clear which boundary condition has been taken into account. However,
in Appendix $C$, we
have considered a quite general
twisted boundary condition compatible with the $Osp(1|2)$ invariance and it is
still noted the presence of such phase factor.
Moreover, as it has been discussed in Appendix $C$, the absence of the factor
$-(-1)^r $ indicates an inconsistency with
the expected degeneracy of a $Osp(1|2) $ invariant system. Hence, this forces
us to conclude that the Bethe ansatz equation
without the factor $-(-1)^r $ either  represents  only the odd part of the
spectrum and therefore is incomplete or is
related to some peculiar boundary condition incompatible with the $Osp(1|2)$
symmetry . In any case, the underlying
quantum field theory  should then be
different from that found in section 4, since that such factor is crucial  in
the computation
of the corresponding critical exponents.}
. The spin chain (8) commutes with the $U(1)$ charge and the index $r$ labels
the
disjoint sectors of the theory with magnetization $r =\sum_{i}^{L} S_i^z $. The
eigenenergies $E^r(L)$ of
such Hamiltonian in a given sector $r$ is obtained by taking the logarithmic
derivative of
$ \Lambda(\lambda, \{\lambda_j \})$ at $\lambda=0$, namely
\EQ
E^r(L) =-\sum_{j=1}^{L-r} \frac{1}{\lambda_j^2 +1/4} + L
\EN

We believe that an important feature of equation (15) is the presence of the
phase factor $(-1)^r $ distinguishing the behaviour
of solutions $\{ \lambda_j \} $ in the odd and even sectors of the theory.
Hence, the bare phase-shift at equal
rapidities $\lambda_j = \lambda_k $ can assume both positive and negative
values, depending on the sector $r$.
These signs may be connected \cite{AL} to the different possibilities
of exchanging ( statistical behaviour) the excitations ( periodic/antiperiodic
boundary conditions of the wave function) in the model. The physical
consequence of this fact is the
evidence of an explicit separation between the fermionic and bosonic degrees of
freedom of the system. These observations
strongly indicate the presence of an extra symmetry in the spin-1 chain (8). In
fact, the $Osp(1|2)$ chain (besides
the $U(1)$ charge ) commutes with the even generators generated by
$\tilde{S}^{\pm}$ (see Appendix A). The spin-1
version of this invariance
is the following commutation relation
\EQ
[H, \sum_{i=1}^{L} (S_i^{\pm})^2]=0
\EN
The conserved charge (17) implies that sectors differing of step $\pm 2$ can
share common eigenvalues. Such symmetry
will be extremely important in the characterization of the finite size
properties to be discussed in the next
sections.

\section{ The completeness of the Bethe ansatz for L=4 sites}

This section is concerned with the completeness of the Bethe ansatz equation
(15) for $L=4$ and consequently
with the numerical study of the behaviour of roots $\{\lambda_j \}$ for a
$finite$
number of sites. This analysis may lead us to discover interesting structures
of roots $\{ \lambda_j \} $ and
also to verify whether or not these
solutions are complete. Such study is motivated by the appearance of a peculiar
bare phase shift ( right hand side of (15) ) in the Bethe ansatz equations. In
fact, at first glance
 one already notices that, in the
even sector, the phase $-(-1)^r$\footnote{
We notice that similar phase sign has previously appeared in the $SL(1|1)$
model ( spin-1/2 XX chain), see e.g. ref. \cite{DV}.} prohibits
symmetric solutions $\{ \lambda_j \} $ possessing one of the roots at the
origin.

In Table 1, for $L=4$, we present the possible configurations of zeros $ \{
\lambda_j \} $ and their corresponding
eigenvalues of energy and momenta. This has been done by numerically solving
the Bethe ansatz
equations (15) and comparing them to exact diagonalization of the $Osp(1|2)$
chain. In our notation,
$\{ n\}$ refers to the possible standard $n$-string \footnote{ We recall that a
$n$-string is characterized by the
root $\lambda_j^{n,\alpha} = \chi_j^n +\frac{i}{2}( n+1 -2 \alpha) $
$\alpha=1,2, \cdots, n $; where $\chi_j$ is a
real number and $n$ is the length of the string.} structures appearing as a
solution of (15). The subscript
$k$ in $n_k$ is the integer or half-integer number which better represents the
logarithmic branches of
equation (15) for the real part of solution $\{ \lambda_j \} $ (see e.g.
equation (18) ).
The first consequence of this study is to show that the Bethe ansatz solutions
produce the complete spectrum
of the $Osp(1|2)$ spin chain, at least for $L=4$. Here, we  have evidently
taken advantage of the hidden symmetry
discussed in section 2 (equation 17), by separating
the even and odd sectors of the theory. Previous experience with
other spin chains \cite{MA1} would suggest that these results are a strong
indication of the
completeness of the roots $ \{ \lambda_j \} $ even in the thermodynamic limit.

In Table 2 we explicitly present some of the roots $\{ \lambda_j \} $ in order
to exemplify our notation
of Table 1. As it has already been noticed, we have observed that the even
sector does not admit a root (symmetric) exact
on the origin. Instead, they prefer to form anomalous string configurations
which have been characterized by the
symbols $A$ and $B$. The first structure $A$ involves four zeros and one may
think that in the thermodynamic
limit they would lead to a 3-string plus one 1-string at the origin. By solving
the Bethe ansatz equations for
several values of $L >4 $, we have verified
that this is possible, but corresponds to a higher excited state on the
spectrum. On the
contrary,  a lower excitation is produced
when the smaller imaginary part of structure $A$ grows and its bigger
imaginary part decreases.
This effect is shown in Table 3 for lattices up to 12 sites \footnote{ For a
quantitative analysis of the
finite size properties one has to go beyond $L > 12 $ in order to take into
account possible logarithmic
corrections . }.
Definitely, this anomaly can not be understood in terms of the usual string
formulation. Analogously, we have
verified that the structure $B$ does not go to $\pm i 1.5$ and thus, together
with $2_0^{\dagger}$,
 forming a 4-string. Again,
for the lowest excitation the imaginary part of $B$ decreases.

In general, we have verified that for higher $L$ the root system is in fact
plagued with such anomalous
string structures. Moreover, in the course of our Bethe ansatz computations we
have noticed an interesting
resemblance to certain properties of the $O(3)$ invariant spin chain. Some
years ago, the author \cite{MA2} has shown that
fractional strings do appear in this model. This is an indication that the
usual string hypothesis has to
be modified for the $Osp(1|2)$ chain. So far a precise reformulation of this
hypothesis has
eluded us. However,
we hope to return to this matter, since we believe that this new structure
will play an important role in
the thermodynamic properties.

\section{ The thermodynamic limit and the finite-size behaviour}
We start this section by investigating the thermodynamic limit of the Bethe
ansatz equation (15). We shall
concentrate our analysis in the ground state of a given sector $r$. In this
case the solutions $\{ \lambda_j \}$ are
real \footnote{
This structure has been determined by  numerically solving the Bethe ansatz
equations (15)
for many values of the lattice size $L$.}
 and by taking the logarithmic of equation (15) we find
\EQ
L \psi_{1/2}(\lambda_j) =2 \pi Q_j + \sum_{k=1,k \neq j}^{L-r}[
\psi_{1}(\lambda_j-\lambda_k) -
\psi_{1/2}(\lambda_j-\lambda_k)]
\EN
where $\psi_a(x)= 2\arctan(x/a)$ and $Q_j$ are integer or semi-integer numbers
defining the different
branches of the logarithm. For those states we find
\EQ
Q_j= -\frac{[L-r-1]}{2} +j-1,~~ j=1,2, \cdots, L-r
\EN

For large $L$, the roots tend toward a continuous distribution with density
$\rho_L^r(\lambda) $ given by
\EQ
\rho_L^r(\lambda) = \frac{d}{d \lambda}Z_L^r(\lambda)
\EN
where the counting function $Z_L^r(\lambda)$ \cite{DW} is defined by
\EQ
Z_L^r(\lambda_j)=\frac{Q^r_j}{L} = \frac{1}{2 \pi} \left \{
\psi_{1/2}(\lambda_j) -\frac{1}{L}
\sum_{k=1,k \neq j}^{L-r}[ \psi_1(\lambda_j-\lambda_k
-\psi_{1/2}(\lambda_j-\lambda_k)] \right \}
\EN

Strictly in the thermodynamic limit ($L \rightarrow \infty $), the system
(20,21) goes into an integral
equation for the density $\rho_{\infty}(\lambda)$ given by
\EQ
2 \pi \rho_{\infty}(\lambda) +\int_{-\infty}^{ \infty} [\psi_1^{'}(\lambda
-\mu) -
\psi_{1/2}^{'}(\lambda-\mu)] \rho_{\infty}(\mu) d \mu = \psi_{1/2}^{'}(\lambda)
\EN
where the prime symbol stands for the derivative. This
equation is then solved by using the Fourier transform
method and we find
\EQ
\rho_{\infty}(\lambda) =\frac{2}{\sqrt{3}} \frac{\cosh(2 \pi
\lambda/3)}{\cosh(4 \pi \lambda/3) +1/2}
\EN
and from equation (16) the ground state energy per site $e_{\infty}$ is
calculated to be
\EQ
e_{\infty} = -\int_{-\infty}^{\infty} \frac{\rho_{\infty}(\lambda)}{\lambda^2
+1/4}
d \lambda +1 = -\frac{4 \pi \sqrt{3}}{9} +1 \cong -1.4184...
\EN

Now we turn to the finite size corrections of the lowest energies $E^r(L)$ of a
given sector $r$. Our
computation will be based on a method introduced by De Vega and Woynarovich
\cite{DW} and further developed
in order to be applied for spin chains \cite {WE,QI} and to the Hubbard model
\cite{WE1,KO3}. Using this
approach we are able to write analytical expressions for the difference of the
energies and density of roots from their corresponding bulk values. Following
ref. \cite{DW} we have
\EQ
\frac{E^r(L)}{L} -e_{\infty} =-2 \pi \int_{-\infty}^{\infty} \rho_{\infty}(\mu)
S_L^r(\mu) d \mu
\EN
and
\EQ
\rho_L^r(\lambda) -\rho_{\infty}(\lambda) = -\frac{1}{2 \pi} \int_{-\infty}^{
\infty} p(\lambda-\mu)S_L^r(\mu) d \mu
\EN
where function $S_L^r(\mu)$ and the Fourier transform of function $p(x)$ are
defined by
\EQ
S_L^r(\mu) =\frac{1}{L} \sum_{j=1}^{L-r}[\delta(\lambda_j -\mu)
-\rho_L^{r}(\mu)]
\EN
\EQ
[1-p(\omega)]^{-1} =G_{+}(\omega) G_{-}(\omega) = \frac{e^{-|\omega|/2}
\Gamma(1/2 -i\omega/4\pi)
\Gamma(1/2 +i\omega/4 \pi)} {\Gamma(1/2 -i3 \omega/4 \pi) \Gamma(1/2 +i3
\omega/4 \pi) }
\EN
where the Fourier transform of $p(x)$ is defined as $p(\omega)=\frac{1}{2 \pi}
\int_{-\infty}^{\infty}
e^{ix \omega} p(x) d x $. According to this technique, the first order
corrections can be calculated
with the help of the Euler-Maclarium formula and equations (25,26) can be
rewritten up to order $O(1/L^2)$ as
\EQ
\frac{E^r(L)}{L} -e_{\infty} =4 \pi \left \{ \int_{\Lambda}^{ \infty}
\rho_{\infty}(\lambda) \rho_L^r(\lambda) d \lambda
-\frac{\rho_{\infty}(\Lambda)}{2L} -\frac{\rho_{\infty}^{'}(\Lambda)}{12 L^2
\rho_L^r(\Lambda)} \right \}
\EN
where the density $ \rho_{L}^r(\lambda +\Lambda) $  satisfies the following
Wiener-Hopf integral equation
\EQ
X^r(t) = \rho_{\infty}(\lambda) + \frac{1}{2 \pi} \left \{\int_{0}^{\infty}
X^r(t) p(t-\mu) d \mu -\frac{p(t)}{2 L}
-\frac{p^{'}(t)}{12 L^2} \right \}
\EN
where $t=\lambda - \Lambda$ ,$X^r(\lambda) = \rho_L^r(\lambda +\Lambda) $
and $\Lambda$ is the largest magnitude root
determined by the boundary condition
\EQ
\int_{\Lambda}^{\infty} \rho_L^r(\lambda) d \lambda = \frac{1}{2L} +
\frac{r}{2L}
\EN

This integral equation is solved by introducing the Fourier transform
\EQ
X_{\pm}^r(\omega) = \int_{-\infty}^{\infty} e^{i \omega t} X_{\pm}^{r}(t);~~~
 X_{\pm}^r(t) \left \{ \begin{array}{ll} X^r(t) & \mbox{ $t>_< 0$} \\
                                          0 & \mbox{$t<_> 0$} \end{array}
\right.
\EN
and after some algebra (see e.g. \cite{QI} ) we find
\EQ
X_{+}^r(\omega) = C^r(\omega) +G_{+}(\omega)[ Q_{+}(\omega) +P(\omega) ]
\EN
where
\EQ
C^r(\omega) =\frac{1}{2L} -\frac{i \omega}{12 L^2 \rho_L^r(\omega)}, ~~~
Q_{+}(\omega)= \frac{2}{\sqrt{3}}
\frac{G_{-}(-i2 \pi/3)}{2 \pi/3 -i \omega} e^{- 2 \pi \Lambda/3}
\EN
\EQ
P(\omega) =-\frac{1}{2L} +\frac{ig}{12L^2 \rho_L^r(\Lambda)} -\frac{i
\omega}{12 L^2 \rho_L^r(\Lambda)},~~ g=-\frac{1}{9}
\EN

Finally, using all these results in equation (29) and approximating
$\rho_{\infty}(\Lambda) \simeq \frac{2}{\sqrt{3}}
e^{-2 \pi \Lambda/3} $ we obtain the first correction for the lowest energy
sector as
\EQ
\frac{E^r(L)}{L} -e_{\infty} = \frac{ \pi^2 \xi}{L^2} ( -\frac{1}{6}
+\frac{r^2}{2 G_{+}(0)^2} )
\EN
where  the sound velocity is calculated to be $\xi=2 \pi/3$ \cite{MAR} and
$G_{+}(0)^2 =1$.

Considering the predictions of conformal invariance for a finite system of size
$L$, this last result leads to
a central charge $c=1$. The conformal dimensions $X_r$ associated to the lowest
state on the sector $r$ are
\EQ
X_r =\frac{r^2}{4}
\EN

This operator content has to be understood in the context of a Gaussian model
with a coupling constant proportional
to $1/4$. In general, besides the `` spin-wave '' state $r$, we shall expect
that  the
complete operator content has also a ``vortex '' excitation parametrized by the
index $m$. The
conformal dimensions are then given by
\EQ
X_{r,m} = \frac{r^2}{4} +m^2
\EN

Such conformal dimensions are in accordance with
the hidden symmetry discussed in section 2. For instance, even excitations
on certain sector $r=2n$ have to be included on the zero sector with dimensions
$n^2$. Indeed from (38) we see that $X_{2n,0}=
X_{0,n}$.

\section{ Concluding remarks}

We have shown that the quantum $Osp(1|2)$ spin chain is solvable by the Bethe
ansatz approach. In particular,
we find that the Bethe ansatz equations present a new property of explicitly
distinguishing the even and odd sectors
of the theory. This feature, as discussed in Appendix A, is a direct
consequence of an extra symmetry. Remarkably
enough, such symmetry resembles much that of a
multiplicative fermionic parity appearing in  supersymmetric field theories.
Analogously, this index can project out the fermionic and bosonic parts of the
superfield. In our model, this
invariance has an important influence on the finite size effects, even inducing
the appearance of new kind
of string structures.

After an appropriate reformulation, the $Osp(1|2)$
spin chain can be seen as certain limit of the Izergin-Korepin \cite{IK,JI,JI1}
vertex model. As a consequence, we are able to show that the Izergin-Korepin
system admits an extra isotropic
solution besides the known $SU(3)$ invariant point \cite{RE}. Due to this novel
property, we believe that our
formulation of the Izergin-Korepin model presented in Appendix B is the most
natural
one for studying the bosonic/fermionic splitting in  a reduced $c=1$ conformal
field theory. The simplest reduction
should be the tricritical Ising model which present a Neveu-Schwartz and Ramond
sectors of excitation \cite{TIM}. Hopefully, the
study of our spin chain with both periodic and antiperiodic boundary conditions
will be able to select  the even
and odd sectors of the theory. Work on this direction is in progress.

The results of this paper also suggest to look for a more general spin
Hamiltonian possessing the $Osp(1|2)$  invariance.
One possibility is as follows
\EQ
H = \sum_{i=1}^{L} \{ J_1 C_{i,i+1} +J_2 C_{i,i+1}^2 \}
\EN
where $J_1,J_2$ are free parameters and $C_{i,i+1}$ is the $Osp(1|2)$ Casimir
operator (see Appendix A). This theory
has at least three integrable points. The point $J_2/J_1= 5/9$
is the critical $Osp(1|2)$ chain. At $J_2/J_1=1/3$ the
Hamiltonian is proportional to the
graded permutation operator, thus possessing an $Sl(1|2)$ symmetry
\cite{CLS,STR}. The ground state
is ferromagnetic and the excitations are gapless. The third point is $J_1=0$
and the model is just the $Osp(1|2)$
Temperely-Lieb operator \cite{MA}. In this case the Temperely-Lieb parameter is
negative and the model is still massless.
Indeed one can show its correspondence with an appropriate point of the
deformed biquadratic spin-1 chain \cite{MA}. Therefore,
in contrast to the bilinear biquadratic $SU(2)$ invariant spin-1 chain (see
e.g. ref. \cite{KU} ), all integrable points
seem to present only massless degrees of freedom. It should be an interesting
problem to discuss the phase
diagram of the Hamiltonian (39) and in particular to verify whether or not
there exists massive regimes.

Finally, we recall that it is possible to construct more  general $R$-matrices
presenting a $Osp(n|2m)$ invariance \cite{BA,CH,MA,OSP,ITO,MS}
. Then, one would like to ask if the rich feature found in the spectrum of the
$Osp(1|2)$ chain
can be even more general for these other systems.
 Our recent results on the Bethe ansatz for the $Osp(1|2n)$ chain \cite{MA3}
and that of ref \cite{MS} for the
$Osp(2|2)$ model strongly indicate that interesting properties are still to be
discovered.

\section*{Acknowledgements}
It is a pleasure to thank M. Malvezzi for his help with numerical checks and
with the figure. We thank F.C. Alcaraz for
innumerable discussions. We also thank one of the referees for pointing us
ref.\cite{KUU}. This work is
supported by CNPq and FAPESP (Brazilian agencies).

\vspace*{1.0cm}
\centerline{\bf Appendix A}
\setcounter{equation}{0}
\renewcommand{\theequation}{A.\arabic{equation}}
The purpose of this appendix is to
write the Hamiltonian (3) in terms of the Casimir operator of the $Osp(1|2)$
algebra.
Such operator has the following expression \cite{RI}
\EQ
C_{i,i+1}= 4 {\tilde{S}}_i^z \stackrel{s}{\otimes} {\tilde{S}}_{i+1}^z +2
[{\tilde{S}}_i^+ \stackrel{s}{\otimes} {\tilde{S}}_{i+1}^-
+{\tilde{S}}_i^- \stackrel{s}{\otimes} {\tilde{S}}_{i+1}^+ ] +
4[V_i^+ \stackrel{s}{\otimes} V_{i+1}^-
-V_i^- \stackrel{s}{\otimes} V_{i+1}^+]
\EN
where the even (bosonic) generators ${\tilde{S}}^{\pm},{\tilde{S}}^z$ are
\begin{eqnarray}
{\tilde{S}}^z =\left( \begin{array}{ccc}
	1/2 & 0 & 0 \\
	0 & 0 & 0  \\
	0 & 0 & 1/2 \\
	\end{array}
	\right) \nonumber &
{\tilde{S}}^+=({\tilde{S}}^-)^t =\left( \begin{array}{ccc}
	0 & 0 & 1 \\
	0 & 0 & 0  \\
	0 & 0 & 0 \\
	\end{array}
	\right) \\
\end{eqnarray}
and the odd (fermionic) operators $V^{\pm}$ are
\begin{eqnarray}
{V}^+ =\left( \begin{array}{ccc}
	0 & 1/2 & 0 \\
	0 & 0 & 1/2  \\
	0 & 0 & 0 \\
	\end{array}
	\right) \nonumber &
V^-=\left( \begin{array}{ccc}
	0 & 0 & 0 \\
	-1/2 & 0 & 0  \\
	0 & 1/2 & 0 \\
	\end{array}
	\right) \\
\end{eqnarray}

In equation (A.1) the  symbol $\stackrel{s}{\otimes}$ stands for the
supertensor product between two matrices. More
precisely we have
\EQ
(A \stackrel{s}{\otimes} B)_{ab}^{ij} = (-1)^{p(i)p(j) +p(a)p(b) +p(i)p(B)}
A_{ai} B_{bj}
\EN
where $p(f)$ is the Grassmann parity of the object $f$ ( vector index or
matrix). In our case we have the
graduation $FBF$ for the space index $i=1,2,3$,
$p(\tilde{S}^{\pm})=p(\tilde{S}^z)=0$ and $p(V^{\pm})=1$.  Using equations
(A.1-3) and the
latest definitions, the Casimir operator is rewritten  as a $9 \time 9$ matrix
of form
\EQ
C_{i,i+1}=\left( \begin{array}{ccccccccc}
	1 &   0 & 0 & 0 & 0 & 0 & 0 & 0 & 0 \\
	0 &   0 & 0 & -1 & 0 & 0 & 0 & 0 & 0 \\
	0 &   0 & -1 & 0 & -1 & 0 & 2 & 0 & 0 \\
	0 &   -1 & 0 & 0 & 0 & 0 & 0 & 0 & 0 \\
	0 &   0 & 1 & 0 & 0 & 0 & -1 & 0 & 0 \\
	0 &   0 & 0 & 0 & 0 & 0 & 0 & -1 & 0 \\
	0 &   0 & 2 & 0 & 1 & 0 & -1 & 0 & 0 \\
	0 &   0 & 0 & 0 & 0 & -1 & 0 & 0 & 0 \\
	0 &   0 & 0 & 0 & 0 & 0 & 0 & 0 & 1 \\
	\end{array}
	\right)
\EN

In order to make the connection between this operator and the Hamiltonian (3)
one has to redefine the grading to $BFF$. After
this canonical transformation we find the fundamental relation
\EQ
C_{i,i+1} =E^g_{i,i+1} -P^g_{i,+1}
\EN

Finally, by using  equation (A.6) we can write the spin chain as
\EQ
H=\sum_{i=1}^{L} \{ C_{i,i+1} +\frac{5}{9} (C_{i,i+1})^2 -\frac{5}{9} \}
\EN
where we have used the important braid-monoid properties $E_g^2 = -E_g$ and
$P_g E_g = E_g P_g$ proved in ref. \cite{MA}.
{}From all these results it is possible to find the  following commutation
relations
\EQ
[H, \sum_{i=1}^{L} \tilde{S}_i^z]=0
\EN
\EQ
[H, \sum_{i=1}^{L} \tilde{S}_i^{\pm}]=0
\EN
\vspace{1.0cm}\\
\centerline{\bf Appendix B}
\setcounter{equation}{0}
\renewcommand{\theequation}{B.\arabic{equation}}
Here we first discuss the relation between the $Osp(1|2)$ and the
Izergin-Korepin vertex model
\cite{IK,JI,JI1}. In order to do that
we have to reformulate conveniently the Boltzmann weights of the IK system. The
$R$-matrix in which we are interested
can be written as
\EQ
R(\lambda)=\left( \begin{array}{ccccccccc}
	a(\lambda) &   0 & 0 & 0 & 0 & 0 & 0 & 0 & 0 \\
	0 &   b(\lambda) & 0 & 1 & 0 & 0 & 0 & 0 & 0 \\
	0 &   0 & c(\lambda) & 0 & e^+(\lambda) & 0 & f^+(\lambda) & 0 & 0 \\
	0 &   1 & 0 & b(\lambda) & 0 & 0 & 0 & 0 & 0 \\
	0 &   0 & e^-(\lambda) & 0 & g(\lambda) & 0 & e^+(\lambda) & 0 & 0 \\
	0 &   0 & 0 & 0 & 0 & b(\lambda) & 0 & 1 & 0 \\
	0 &   0 & f^-(\lambda) & 0 & e^-(\lambda) & 0 & c(\lambda) & 0 & 0 \\
	0 &   0 & 0 & 0 & 0 & 1 & 0 & b(\lambda) & 0 \\
	0 &   0 & 0 & 0 & 0 & 0 & 0 & 0 & a(\lambda) \\
	\end{array}
	\right)
\EN
where
\EQ
a(\lambda) =\frac{
\sin[\gamma(i\lambda+1)]}{\sin[\gamma]},~~b(\lambda)=-\frac{\sin[i\gamma
\lambda]}{\sin[\gamma]},~~
c(\lambda)=\frac{\sin[i \gamma \lambda] \sin[\gamma(i \lambda
+1/2)]}{\sin[\gamma] \sin[\gamma(i \lambda +3/2)]}
\EN
and
\EQ
e^{\pm}= \mp i e^{\mp i \gamma} \frac{\sin[i \gamma \lambda]}{\sin[\gamma(i
\lambda +3/2)]},
{}~ f^{\pm}=a(\lambda) -
e^{\mp i 2 \gamma} c(\lambda),~~ g(\lambda)= -\frac{\sin[i \gamma
\lambda]}{\sin[\gamma]} +\frac{\sin[3 \gamma/2]}
{\sin[\gamma(i \lambda +3/2)]}
\EN

It is easy to check that this solution satisfies the properties appearing in
the IK vertex model \cite{IK,RE,TA}. For instance,
to recover the original notation of ref \cite{TA} one has to roughly replace
$\lambda/2 \rightarrow \lambda$ and
$\eta \rightarrow \pi/2 +i \gamma/2$.
We also have to use some symmetries of this solution, such as $b \rightarrow
-b$ and
$e^{\pm} \rightarrow \pm i e^{\pm}$. The Hamiltonian associated with the vertex
model (B.1) is calculated to be
\bear
H= \tilde{J} \sum_{i=1}^{L} \left[ \sin(\gamma/2) \sigma_i^2
-\sin(3 \gamma/2) \sigma_i -2\sin(\gamma) \cos(3\gamma/2)
[(\sigma_i^z)^2 - \sigma_i^z]
\right . \nonumber \\ \left .
-2\sin(\gamma) \cos(\gamma/2)(1-\sin(\gamma/2))[\sigma_i^z \sigma_i^{\perp}
+ \sigma_i^{\perp} \sigma_i^z]
+2\sin(3 \gamma/2) \cos^2(\gamma/2) [(S_i^z)^2 +(S_{i+1}^z)^2]
\right . \nonumber \\ \left .
-[\sin(\gamma/2) +2\cos^2(\gamma/2) \sin(3 \gamma/2)]I +
\frac{i}{2} \sin(\gamma/2) \sin(2 \gamma)[ (S_i^z)^2 S_{i+1}^z - S_i^z
(S_{i+1}^z)^2 ]
\right . \nonumber \\ \left .
-i \frac{\sin(2 \gamma)}{4}[(S_{i+1}^z-S_{i})\sigma_i^{\perp}
+\sigma_i^{\perp}(S_{i+1}^z-S_i^z)
\right ]
\ear
where $\tilde{J} =J \frac{\gamma}{\sin(\gamma) \sin(3 \gamma/2)} $.
Such Hamiltonian can be diagonalized by using the analytical approach
of section 2 or by directly applying the algebraic calculation of
ref \cite{TA}. As a final result we find that the spectrum is parametrized
by the following Bethe ansatz equations
\EQ
{\left( \frac{\sinh[\gamma(\lambda_j -i/2)]}{
\sinh[\gamma(\lambda_j +i/2)]} \right)}^{L}= - (-1)^r \prod_{k=1}^{M}
 \frac{\sinh[\gamma(\lambda_j -\lambda_k -i)]}{
\sinh[\gamma(\lambda_j-\lambda_k +i)]}
\frac{\sinh[\gamma(\lambda_j -\lambda_k +i/2)]}{
\sinh[\gamma(\lambda_j-\lambda_k -i/2)]}
\EN
where the eigenenergies are given by
\EQ
E^r(L)= \sum_{i=1}^{L-r} \frac{2 \gamma \sin(\gamma)}{\cos(\gamma) -\cosh(
2 \gamma \lambda_j)} + \frac{\gamma \cos(\gamma)}{\sin(\gamma)}
\EN

Hence, taking the limit $ \gamma \rightarrow 0 $ and collecting factors
up to first order in $\gamma$ we then recover the isotropic spin-1  Hamiltonian
(8).
It is crucial to notice that the last term of equation (B.4) and
(8) are
equivalent by a canonical transformation of type $e^{\pm} \rightarrow \pm i
e^{\pm}$. Moreover, from
equations (B.5, B.6) the  limit $ \gamma \rightarrow 0 $ directly recovers the
Bethe
equations
of $Osp(1|2)$ chain.

At this point we recall that the quantum spin chain associated with the
$IK$ vertex model has been recently discussed in the literature ( see e.g.
\cite{WA,VL} and references therein). However, in all cases the
problem has been formulated in such way that one obtains the $SU(3)$ invariant
chain as the isotropic point. Remarkably enough, this other isotropic branch
limit
,the $Osp(1|2)$ chain discussed in this paper, has not been noticed
before\footnote{
For instance, considering notation of ref \cite{WA} one must do the shift
$\gamma \rightarrow
\pi-\gamma$ before making the limit $ \gamma \rightarrow 0$.}. Thus
, in our formulation the $IK$ model is seen as a certain deformation of the
$Osp(1|2)$ chain.
We believe that this is  the most appropriate formulation of the problem, in
order to get reduced conformal theories presenting both fermionic and
bosonic degrees of freedom.

Now we are going to summarize the solution of the $Osp(1|2)$ chain by the
algebraic Bethe
ansatz approach developed by Tarasov \cite{TA}. In this case one works directly
with the
operator content of the monodromy matrix $\tau({\lambda})$
\EQ
\tau(\lambda) =
 {\cal L}_{0L}(\lambda) \cdots {\cal L}_{01}(\lambda)
=\left( \begin{array}{ccc}
	A_1(\lambda) &   B_1(\lambda) & B_2(\lambda) \\
	C_1(\lambda) &  A_2(\lambda)& B_3(\lambda) \\
	C_2(\lambda) & C_3(\lambda) & A_3(\lambda) \\
	\end{array}
	\right)
\EN
and with the integrability condition
\EQ
\tilde{{\cal L}}(\lambda -\mu) \tau(\lambda) \otimes \tau(\mu)
=\tau(\mu) \otimes \tau(\lambda) \tilde{{\cal L}}( \lambda -\mu)
\EN
where $\tilde{{ \cal L}}_{ab}^{cd}(\lambda) = {\cal L}_{ba}^{cd}(\lambda) $.

On the reference state $ |0> $ we have
\EQ
T(\lambda) |0> = \sum_{i=1}^{3} A_i |0> = (i-\lambda)^{L} +\lambda^{L} + \left
[ \frac{\lambda(i/2
-\lambda)}{(\lambda -3i/2)} \right ]^{L}
\EN

The one particle excitation $|\phi(\lambda_1) >$ over the pseudovacuum is given
by
\EQ
| \phi(\lambda_1) > = B_1(\lambda_1) |0>
\EN

In order to calculate the Transfer matrix acting on the one particle state one
has to take
advantage of the following commutation relations coming from equation (B.8)
\EQ
A_1(\lambda) B_1(\lambda_1) |0> = f_1(\lambda_1 -\lambda) B_1(\lambda_1)
A_1(\lambda) |0> -
f_2(\lambda_1 -\lambda) B_1(\lambda) A_1(\lambda_1) |0>
\EN
\bear
A_2(\lambda) B_1(\lambda_1) |0> = \frac{f_1(\lambda-
\lambda_1)}{f_3(\lambda-\lambda_1)}
B_1(\lambda_1) A_2(\lambda) |0> + f_2(\lambda_1 -\lambda) B_1(\lambda)
A_2(\lambda_1) |0> \nonumber \\
+f_4(\lambda-\lambda_1) B_3(\lambda) A_1(\lambda_1) |0>
\ear
\EQ
A_3(\lambda) B_2(\lambda_1) |0> = f_5(\lambda - \lambda_1) B_1(\lambda_1)
A_3(\lambda) |0>
-f_4(\lambda- \lambda_1) B_3(\lambda) A_2(\lambda_1) |0>
\EN
where
\EQ
f_1(x) = \frac{i -x}{x};~~ f_2(x) = \frac{i}{x};~~ f_3(x) =\frac{i/2 -x}{i/2
+x};~~f_4(x)=\frac{i}{i/2-x};~
f_5(x)= \frac{x -3i/2}{i/2 -x}
\EN

By using these relations we find that
\EQ
T(\lambda)| \phi(\lambda_1)> = \sum_{i=1}^{3} A_i(\lambda) B_1(\lambda_1) |0> =
\Lambda(\lambda,\lambda_1)
|\phi(\lambda_1) >
\EN
where
\EQ
\Lambda(\lambda,\lambda_1) = (i-\lambda)^{L} \frac{i- \lambda_1
+\lambda}{\lambda_1 -\lambda}
\lambda^{L} \frac{i -\lambda +\lambda_1}{\lambda -\lambda_1} \frac{i/2
-\lambda_1 +\lambda}{i/2 +\lambda_1
-\lambda} +
\left [ \frac{ \lambda (i/2 -\lambda)}{ \lambda -3i/2} \right ]^{L} \frac{ 3i/2
-\lambda +\lambda_1}
{\lambda -\lambda_1 -i/2}
\EN
provided that
\EQ
\left ( \frac{i -\lambda_1}{\lambda_1} \right )^{L} =1
\EN

The two particle state is given by the ansatz \cite{TA}
\EQ
|\phi(\lambda_1,\lambda_2)> = [B_1(\lambda_1) B_2(\lambda_2) +\eta(\lambda_1,
\lambda_2) B_2(\lambda_1)
A_1(\lambda_2)]|0>
\EN
where function $\eta(\lambda_1,\lambda_2)$ is determined by imposing that under
permutation
$ \lambda_1 \leftrightarrow \lambda_2$ the symmetric state
$|\phi(\lambda_2,\lambda_1)> $ is at most
proportional to that of equation (B.18) . This is solved by using the
commutation relation
\EQ
B_1(\lambda_2)  B_1(\lambda_1)  = f_3(\lambda_1 -\lambda_2)[B_1(\lambda_1)
B_1(\lambda_2) -
f_4(\lambda_1 -\lambda_2) B_2(\lambda_1) A_1(\lambda_2) ]
+f_4(\lambda_2-\lambda_1) B_2(\lambda_2) A_1(\lambda_1)
\EN
and we find that $\eta(\lambda_1,\lambda_2) =f_4(\lambda_1 - \lambda_2) $.
Moreover, the Transfer matrix
on this symmetrized state has the following eigenvalue
\bear
\Lambda(\lambda,\lambda_1,\lambda_2) =
(i-\lambda)^{L} \prod_{j=1}^{2}\frac{i- \lambda_j +\lambda}{\lambda_j -\lambda}
+\lambda^{L} \prod_{j=1}^{2}\frac{i -\lambda +\lambda_j}{\lambda -\lambda_j}
\frac{i/2 -\lambda_j +\lambda}{i/2 +\lambda_j
-\lambda} +
 \nonumber  \\
\left [ \frac{ \lambda (i/2 -\lambda)}{ \lambda -3i/2} \right ]^{L}
\prod_{j=1}^{2}\frac{ 3i/2 -\lambda +\lambda_j}
{\lambda -\lambda_j -i/2}
\ear
provided that
\EQ
\left ( \frac{i-\lambda_j}{\lambda_j} \right )^{L} = - \prod_{k=1}^{2}
\frac{\lambda_j -\lambda_k -i}{\lambda_j -\lambda_k +i}
\frac{\lambda_j -\lambda_k +i/2}{\lambda_j -\lambda_k -i/2}
\EN

Following the arguments of ref. \cite{TA} these last results can be generalized
for any $n$-particle excitation.
We then recover equation (15) by making the shift $\lambda_j \rightarrow
\lambda_j +i/2 $.

\vspace*{1.0cm}
\centerline{\bf Appendix C}
\setcounter{equation}{0}
\renewcommand{\theequation}{C.\arabic{equation}}
This appendix is concerned with the study of the $Osp(1|2)$ chain with twisted
boundary condition. It is possible to
check that the $Osp(1|2)$ algebraic structure is invariant under the following
twisted transformation :
\EQ
V^{\pm} \rightarrow e^{\pm i \phi/2} V^{\pm},~~
{\tilde{S}}^{\pm} \rightarrow e^{\pm i \phi} {\tilde{S}}^{\pm},~~
{\tilde{S}}^{z} \rightarrow  {\tilde{S}}^{z}
\EN

This means that one can define a more general boundary condition (preserving
the $Osp(1|2)$ algebra) by
\EQ
V_{L+1}^{\pm} \rightarrow e^{\pm i \phi/2} V_1^{\pm},~~
{\tilde{S}}_{L+1}^{\pm} \rightarrow e^{\pm i \phi} {\tilde{S}}_1^{\pm},~~
{\tilde{S}}_{L+1}^{z} \rightarrow  {\tilde{S}}_1^{z}
\EN

At this point it is interesting to remark that, besides the usual periodic case
($\phi=0$), the condition $(C.2)$
admits the interesting mixed case of a periodic (antiperiodic) boundary
condition in the bosonic (fermionic) degrees of freedom
by taking the angle $\phi =2 \pi$. By imposing the boundary $(C.2)$ and from
Appendix $A$ we find that the boundary term
$H_b(\phi)$ of the corresponding $Osp(1|2)$ Hamiltonian is given by
\EQ
H_{b}(\phi)=\left( \begin{array}{ccccccccc}
	1 &   0 & 0 & 0 & 0 & 0 & 0 & 0 & 0 \\
	0 &   0 & 0 & -e^{-i \phi/2} & 0 & 0 & 0 & 0 & 0 \\
	0 &   0 & 2/3 & 0 & \frac{2}{3} e^{-i \phi/2} & 0 & \frac{1}{3} e^{-i \phi} &
0 & 0 \\
	0 &   -e^{i \phi/2} & 0 & 0 & 0 & 0 & 0 & 0 & 0 \\
	0 &   0 & -\frac{2}{3}e^{i \phi/2} & 0 & -5/3 & 0 & \frac{2}{3} e^{-i \phi/2}
& 0 & 0 \\
	0 &   0 & 0 & 0 & 0 & 0 & 0 & -e^{-i \phi/2} & 0 \\
	0 &   0 & \frac{1}{3}e^{i \phi} & 0 & -\frac{2}{3}e^{i \phi/2} & 0 & 2/3 & 0 &
0 \\
	0 &   0 & 0 & 0 & 0 & -e^{i \phi/2} & 0 & 0 & 0 \\
	0 &   0 & 0 & 0 & 0 & 0 & 0 & 0 & 1 \\
	\end{array}
	\right)
\EN

Moreover, by performing the Bethe ansatz analysis of Appendix $B$ we find that
the associated Bethe ansatz equation of the
$Osp(1|2)$ chain with twisted boundary condition $(C.2)$ is
\EQ
{\left( \frac{\lambda_j -i/2}{\lambda_j +i/2} \right)}^{L}= - (-1)^r e^{i
\phi/2} \prod_{k=1}^{M}
\left(\frac{\lambda_j -\lambda_k -i}{\lambda_j-\lambda_k +i} \right)
\left(\frac{\lambda_j -\lambda_k +i/2}{\lambda_j-\lambda_k -i/2} \right)
\EN
and the energy equation (16) remains unchanged.

The importance of this analisis is that it allows us to study the behaviour of
the spectrum
of a theory possessing Bethe ansatz equation (15) without the presence of the
factor $-(-1)^r$.
This is as follows. For odd sectors $-(-1)^r =+ 1$, therefore the behaviour of
the spectrum is precisely
the same found for the $Osp(1|2)$ with periodic boundary condition. For even
sectors, however, this can be studied
by varying adiabaticaly the angle $\phi$ up to $\phi = 2 \pi$. For instance, in
Table 4, we present the ground state
structure of roots $\lambda_j$ for $L=2$ and $r=0$. We observe that when $\phi
\rightarrow 2 \pi$ one of the roots
diverges and the other goes to zero. Moreover, when $ \phi \rightarrow 2 \pi $,
we have numerically verified for several values of $L$ that the usual set of
$L$ roots $\{ \lambda_1(\phi), \cdots, \lambda_L(\phi) \} $ of the ground state
of sector $r=0$
 goes to $\{ \infty, 0, \lambda_3, \cdots
, \lambda_L \}$ where the set $\{ 0, \lambda_3, \cdots, \lambda_L \} $ is
precisely the roots
characterising the ground state of sector $r=1$.
This means that in absence of the phase factor $-(-1)^r $ ( $\phi =2 \pi$ and
$r=0$ ) the ground state of the sectors $r=0$ and $r=1$ are degenerated, since
the root $\lambda = \infty $ gives a null
contribution to the energy. Of course the same reasoning can be repeated for
any pair of sectors
$r=2n$ and $r=2n+1$, $n=0,1,\cdots $ . As a consequence, we conclude that (in
absence of the factor $-(-1)^r $)
the degeneracy is of step $\pm 1$ which is in contradiction to that found (
step $\pm 2$) for a $Osp(1|2)$ invariant
system ( see Appendix $A$, Eq.(A.9)). Nevertheless, due to this degeneracy,
some conformal dimensions found
in section 4 are now prohibited changing remarkably the underlying quantum
field theory. Finally,
we recall that the presence of the factor $-(-1)^r$ changes the topology
of the Bethe ansatz equation for sector $r=1$ and  no such contradiction is
found. Hopefully, this analysis will lend
further support to the physical meaning and to the importance of the phase
factor $-(-1)^r$ in the Bethe ansatz equation.
\vspace{0.4cm}\\

\vspace*{0.5cm}
\newpage

\centerline{ \bf Tables }
\vspace*{0.5cm}
Table 1.(a,b)  The complete set of states for the $Osp(1|2)$ Hamiltonian
for $L=4$. The classification is given in terms of the zeros $\lambda_j$
of the Bethe ansatz equations and their respective energies and momenta.
The states with the superscripts $*$ and $\dagger$
indicate double degeneracy (
${\lambda_j} \rightarrow {- \lambda_j} $) and exact string configuration
,respectively. Some noted special structures
have been denoted by $A$ and $B$.

\begin{center}
{\bf Table ( 1.a)}\\
\vspace{0.5cm}
\begin{tabular}{|l|l|l|l|} \hline
r & $ \{ n \}$ & $ E^r(L)/L$&$  p$ \\ \hline \hline
0 & $1_{3/2} 1_{1/2} 1_{-1/2}1_{-3/2}$ & $-1.487 116$ & $0$\\ \hline
0,2 & $1_{-1/2} 1_{1/2}$ & $-0.795 334$& $ 0$ \\ \hline
0 &  $1_{-3/2} 1_{-1/2} 2_{1/2}$ & $ -0.693 713$ & $\pi/2(3 \pi/2)^{*}
$ \\ \hline
0 & $2_0^{\dagger} 1_{1} 1_{-1} $ & $-0.283 594$ & $\pi$ \\ \hline
0 & $A$ & $-0.279 860$ & $0$ \\ \hline
0,2 & $ 1_{3/2} 1_{-1/2} $ & $ -0.107 625 $ &$ \pi/2(3 \pi/2)^{*}$ \\ \hline
0,2 & $ 1_{3/2} 1_{1/2} $ & $0$ & $\pi$ \\ \hline
0 & $2_{1/2} 2_{-1/2} $ & $ 0.266 976 $ & $0$ \\ \hline
0 & $1_{-3/2} 3_{1/2}$ & $0.360 380$ & $\pi/2(3 \pi/2)^{*}$ \\ \hline
0,2 & $2_0^{\dagger}$ & $0.5$ & $\pi$ \\ \hline
0,2 & $1_{3/2} 1_{-3/2}$ & $ 0.628 667$ & $0$ \\ \hline
0,2 & $ 2_{1/2}$ & $0.774 292$ & $\pi/2(3 \pi/2)^{*} $  \\ \hline
0 & $ 2_0^{\dagger} B $ & $0.783 594 $ & $ \pi$ \\ \hline
0,2,4 & $----$ & $1.0$ & $0$ \\ \hline
\end{tabular}
\end{center}
\newpage
\begin{center}
{\bf Table ( 1.b)}\\
\vspace{0.5cm}
\begin{tabular}{|l|l|l|l|} \hline
r & $ \{ n \}$ & $ E^r(L)/L$&$  p$ \\ \hline \hline
1 & $1_{-1} 1_{0} 1_{-1}$ & $-1.350 519 $ & $\pi$\\ \hline
1 & $1_{0} 1_{1}$ & $-0.567 521$& $ \pi/2(3 \pi/2)^{*}$ \\ \hline
1 &  $1_{0}  2_{1}$ & $ -0.333 33 \stackrel{.}{3} $ & $\pi/2(3 \pi/2)^{*}
$ \\ \hline
1 & $1_{1} 1_{-1} $ & $-0.166 66 \stackrel{.}{6}$ & $0$ \\ \hline
1 & $1_{-1} 2_{1}$ & $0$ & $\pi/2(3\pi/2)^{*}$ \\ \hline
1,3 & $ 1_{0}  $ & $ 0 $ &$ \pi$ \\ \hline
1,3 & $ 1_{1}  $ & $0.5$ & $\pi/2(3 \pi/2)^{*}$ \\ \hline
1 & $2_{0}^{\dagger}  $ & $ 0.5 $ & $\pi$ \\ \hline
1 & $3_{0} $ & $0.638 492 $ & $\pi$ \\ \hline
1 & $2_{1} $ & $0.734 187  $ & $\pi/2(3 \pi/2)^{*}$ \\ \hline
1,3 & $------$ & $1.0$ & $0$ \\ \hline
\end{tabular}
\end{center}
Table 2. Some complex solutions of the Bethe ansatz equations in the sector
$r=0,1$.

\begin{center}
{\bf Table (2)}\\
\vspace{0.5cm}
\begin{tabular}{|l|l|l|l|} \hline
$\{ n \}$ & $ r$ & $ p$&$  \lambda_j$ \\ \hline \hline
$1_{-3/2} 1_{-1/2} 2_{1/2}$ & $0$ & $\pi/2(3 \pi/2)^{*}$ & $-0.681 058; -0.1226
705; 0.403 881 \pm
i0.507 723 $\\ \hline
$A$ & $0$ & $0$& $ \pm i0.097 049; \pm i0.936 067 $ \\ \hline
$2_0^{\dagger} B$ &  $0$ & $\pi $ & $ \pm i/2; \pm i1.418 833
$ \\ \hline
$1_0 2_1$ & $1 $ & $\pi/2(3 \pi/2)^{*}$ & $0.106 997; 0.696 501 \pm i0.464 584$
\\ \hline
$3_0$ & $1$ & $\pi$ & $0, \pm i1.016 413$ \\ \hline
\end{tabular}
\end{center}
\newpage
\vspace{0.5cm}
Table 3.  The imaginary parts of the anomalous solution $A$ for some values of
$L$. $\pm i x$
($\pm i y$ ) is the lowest (biggest) imaginary part of $A$.

\begin{center}
{\bf Table (3)}\\
\vspace{0.5cm}
\begin{tabular}{|l|l|l|l|} \hline
$L$ & $ \pm ix$ & $ \pm iy$&$ E(L)/L $ \\ \hline \hline
$4$ & $\pm i0.097 494$ & $\pm i0.936 0674 $ & $ -0.279 859
$\\ \hline
6 & $\pm i0.103 740$ & $\pm i0.908 041$& $ -0.932 654 $ \\ \hline
$8$&  $\pm i 0.108 785 $ & $\pm i 0.894 917 $ & $-1.152 731$
 \\ \hline
$10$ & $\pm i0.113 551$ & $\pm i0.887 498$ & $-1.251 731$ \\ \hline
$12$ & $\pm i0.117 536 $ & $ \pm i0.882 745 $& $-1.304 385$ \\ \hline
\end{tabular}
\end{center}
\vspace{0.5cm}
Table 4. Structure of the roots of the ground state for $L=2$ and $r=0$ with
twisted boundary condition.
\begin{center}
{\bf Table (4)}\\
\vspace{0.5cm}
\begin{tabular}{|l|l|l|} \hline
$\phi$ & $ \{ \lambda_1, \lambda_2 \}$ & $  E(L)/L $ \\ \hline \hline
$\pi$ & $\{ 0.1160, -0.8216 \} $ &  $ -1.457
$\\ \hline
$\frac{5 \pi}{4}$ & $\{ 0.0701, -1.0355 \} $ &  $ -1.339 $ \\ \hline
$\frac{7 \pi}{10}$ & $\{ 0.0414, -1.3034 \} $ &  $ -1.2425 $ \\ \hline
$\frac{3 \pi}{2}$ & $\{ 0.0313, -1.4593 \} $ &  $ -1.2021 $ \\ \hline
$\frac{5 \pi}{3}$ & $\{ 0.0124, -2.0702 \} $ &  $ -1.1088 $ \\ \hline
$\frac{20 \pi}{11}$ & $\{ 0.0025, -3.6071 \} $ &  $ -1.0376 $ \\ \hline
$2 \pi$ & $\{ 0, \infty \} $ &  $ -1 $ \\ \hline
\end{tabular}
\end{center}
\newpage
\centerline{ \bf Figures }
\vspace*{0.5cm}
Figure 1. The 19 nonvanishing Boltzmann weights of the $Osp(1|2) $ chain. Those
values not explicitly indicated in the
figure are: $a(\lambda,\eta) =-(\lambda -\eta)$ ; $c(\lambda,\eta) =
-\frac{\lambda(\lambda-\eta/2)}{\lambda -3 \eta/2} $;
$e(\lambda,\eta) = \frac{\lambda \eta}{\lambda -3 \eta/2} $ ;
$f(\lambda,\eta) = \eta \frac{2 \lambda -3 \eta/2}{\lambda -3 \eta/2}$ and
$g(\lambda,\eta) =\frac{\lambda(\lambda -3 \eta/2)
-3 \eta^2/2 }{\lambda - 3 \eta/2} $ .

\end{document}